\documentclass[
    ,final            % use final for the camera ready runs
  ]
  {aipproc}

\layoutstyle{8x11single}

\begin{document}

\title{Neutrino interactions with nuclei}


\author{M.~Martini}{
  address={Universit\'e de Lyon, Univ.  Lyon 1, 
 CNRS/IN2P3, IPN Lyon, F-69622 Villeurbanne Cedex}
  ,altaddress={Universit\`a di Bari, I-70126 Bari}
}

\author{G.~Chanfray}{
  address={Universit\'e de Lyon, Univ.  Lyon 1, 
 CNRS/IN2P3, IPN Lyon, F-69622 Villeurbanne Cedex}
}

\author{M.~Ericson}{
  address={Universit\'e de Lyon, Univ.  Lyon 1, 
 CNRS/IN2P3, IPN Lyon, F-69622 Villeurbanne Cedex}
  ,altaddress={Theory division CERN, CH-12111 Geneva}
}

\author{J.~Marteau}{
  address={Universit\'e de Lyon, Univ.  Lyon 1, 
 CNRS/IN2P3, IPN Lyon, F-69622 Villeurbanne Cedex}
}

\begin{abstract}
We present a model for neutrino-nucleus scattering in the energy 
region relevant for present and forthcoming neutrino-oscillation experiments. 
The model is based on the RPA treatment of the nuclear responses in 
the quasi-elastic and Delta-resonance region. It includes also in a phenomenological 
way nucleon knock-out. 
It aims at the description, within a single framework, 
of several final state channels i.e. quasi-elastic, incoherent 
and coherent one-pion production and two- or several-nucleon knock-out.
\end{abstract}

\maketitle

%%%%%%%%%%%%%%%%%%%%%%%%%%%%%%%%%%%%%%%%%%%%
%% MAINMATTER
%%%%%%%%%%%%%%%%%%%%%%%%%%%%%%%%%%%%%%%%%%%%

\section{Introduction}

Various theoretical approaches \cite{various} have been used to interpret 
the experimental results on neutrino interactions 
with matter \cite{Radecky:1981fn,
Nakayama:2004dp,Hasegawa:2005td,:2007ru,AguilarArevalo:2008xs,
:2008eaa,Hiraide:2008eu,AguilarArevalo:2009eb}
in quasi-elastic processes or coherent 
and incoherent single pion production. 

In our work, 
we explore these interactions in the energy 
region around 1 GeV
using the formalism of the nuclear response functions treated in the random 
phase approximation (RPA) and incorporating Delta-resonance excitation 
as in the work of Marteau \cite{Marteau:1999kt}. 
 %Further details will be given in a forthcoming paper \cite{Martini}.
This approach has the merit of describing in a unique frame 
several final state channels i.e. quasi-elastic, incoherent 
and coherent one-pion production and two- or several-nucleon knock-out. 
In the following we present the results obtained for each of them.
%\section{Formalism}

Several types of nuclear responses enter the total neutrino-nucleus cross-section: 
the isovector $R_\tau$, the spin-isospin transverse $R_{\sigma\tau (T)}$ 
or longitudinal $R_{\sigma\tau (L)}$. 
In order to illustrate this point, we give below a simplified expression 
of the double differential cross-section  for the reaction \mbox{$ \nu_l \, (\bar{\nu}_l) + A \longrightarrow l^- \, (l^+) + X $} 
which, in particular, ignores the lepton mass contribution 
and assumes zero $\Delta$ width:
\begin{eqnarray} \label{SIGMANUENU}
\frac{\partial^2\sigma}{\partial\Omega \,\partial k^\prime} & = & \frac{G_F^2 \, 
\cos^2\theta_c \, (\vec{k}^\prime)^2}{2 \, \pi^2} \, \cos^2\frac{\theta}{2} \, 
\left[ G_E^2 \, (\frac{q_\mu^2}{\vec{q}^2})^2 \, R_\tau^{NN} 
%\right. \nonumber \\ 
+  G_A^2 \, \frac{( M_\Delta - M )^2}{2 \, \vec{q}^2} \, R_{\sigma\tau (L)}^{N\Delta} +  G_A^2 \, \frac{( M_\Delta - M )^2}{\vec{q}^2} R_{\sigma\tau (L)}^{\Delta\Delta} \right.\nonumber \\ 
& + & \left( G_M^2 \, \frac{\omega^2}{\vec{q}^2} + G_A^2 \right) \, 
 \left( - \frac{q_\mu^2}{\vec{q}^2} + 2 \tan^2\frac{\theta}{2} \right) \,
\left( R_{\sigma\tau (T)}^{NN} + 2 R_{\sigma\tau (T)}^{N\Delta} 
+ R_{\sigma\tau (T)}^{\Delta\Delta} \right) 
\nonumber \\ 
& \pm & \left. 2 \, G_A \, G_M \, \frac{k + k^\prime}{M} \, 
\tan^2\frac{\theta}{2} \, 
\left( R_{\sigma\tau (T)}^{NN} + 2 R_{\sigma\tau (T)}^{N\Delta} 
+ R_{\sigma\tau (T)}^{\Delta\Delta} \right) \right]. 
\end{eqnarray}
For the variable definitions and for the complete formulas 
we refer to \cite{Marteau:1999kt,Martini}. 
We stress that in the actual 
calculations we make use of the full formulas.

The various responses $R$ appearing in Eq.(\ref{SIGMANUENU}) are
related to the imaginary part of the corresponding 
full polarization propagators through
\begin{equation} \label{eq:6}
R (q,\omega) = -\frac{1}{\pi} \, \mathrm{Im}\Pi(q,q,\omega). 
\end{equation}
They are calculated within a RPA ring approximation, starting 
from ``bare'' propagators (meaning that the nuclear correlations are switched off) 
and solving integral equations which have the generic form

\begin{equation} \label{eq:4}
\Pi = \Pi^0 + \Pi^0 \, V \Pi,
\end{equation}
where $ V $ denotes the effective interaction between \textit{particle-hole} 
excitations.\\ 

The bare polarization propagator is density dependent. 
In a finite system, $\Pi^0(\vec{q},\vec{q}',\omega)$ is non-diagonal in momentum space. 
In order to account for the finite size effects we evaluate it in a semi-classical approximation where it can be cast in the form 
\begin{equation} \label{eq:laktineh}
\Pi^0(\vec{q},\vec{q}',\omega) = \int\, d \vec{r} \,e^{-i(\vec{q}-\vec{q}') \cdot \vec{r}} \,
\Pi^0\left(\frac{\vec{q}+\vec{q}'}{2},\vec{r},\,\omega\right).
\end{equation}
In practice we use a local density approximation, 
\begin{equation}
\Pi^0\left(\frac{\vec{q}+\vec{q}'}{2},\vec{r},\,\omega\right)=
\Pi^0_{k_F(r)}\left(\frac{\vec{q}+\vec{q}'}{2},\,\omega\right), 
\end{equation}
where the local Fermi momentum $ k_F(r) $ is related to the experimental nuclear density through:
\begin{equation}
k_F(r) = ( 3/2 \, \pi^2 \, \rho(r) )^{1/3}.
\end{equation}

The bare response is the sum of the following partial components:       
(1) $ NN $ quasi-elastic (as described by the standard Lindhard function); 
(2) $ NN $  \textit{2p-2h}; 
(3) $ N\Delta $  and $ (3^\prime) $ $ \Delta N $ \textit{2p-2h}; 
(4) $ \Delta\Delta  $ $\pi \, N $;
(5) $ \Delta\Delta $ \textit{2p-2h};
(6) $ \Delta\Delta $ \textit{3p-3h}.

The RPA response generically writes
\begin{equation} 
\mathrm{Im} \Pi = \left| \Pi \right|^2 \, \mathrm{Im} V \,
+  \left| 1 + \Pi \, V \right|^2 \, \mathrm{Im}\Pi^0  \,.  \label{SEPAR}
\end{equation}
It splits in two terms. The first implies a cut on the pion exchange potential $V_\pi$. It
 represents the coherent pion production where the nucleus is left in the ground state. 
The second, proportional to the bare polarization propagator $\mathrm{Im}\Pi^0$, 
reflects the type of final state already mentioned in the bare case, 
modified by collective effects.

\section{Results and comparison with data}

\subsection{Coherent pion production}
\begin{figure}
\label{fig_coh_diff}
\includegraphics[width=0.49\textwidth]{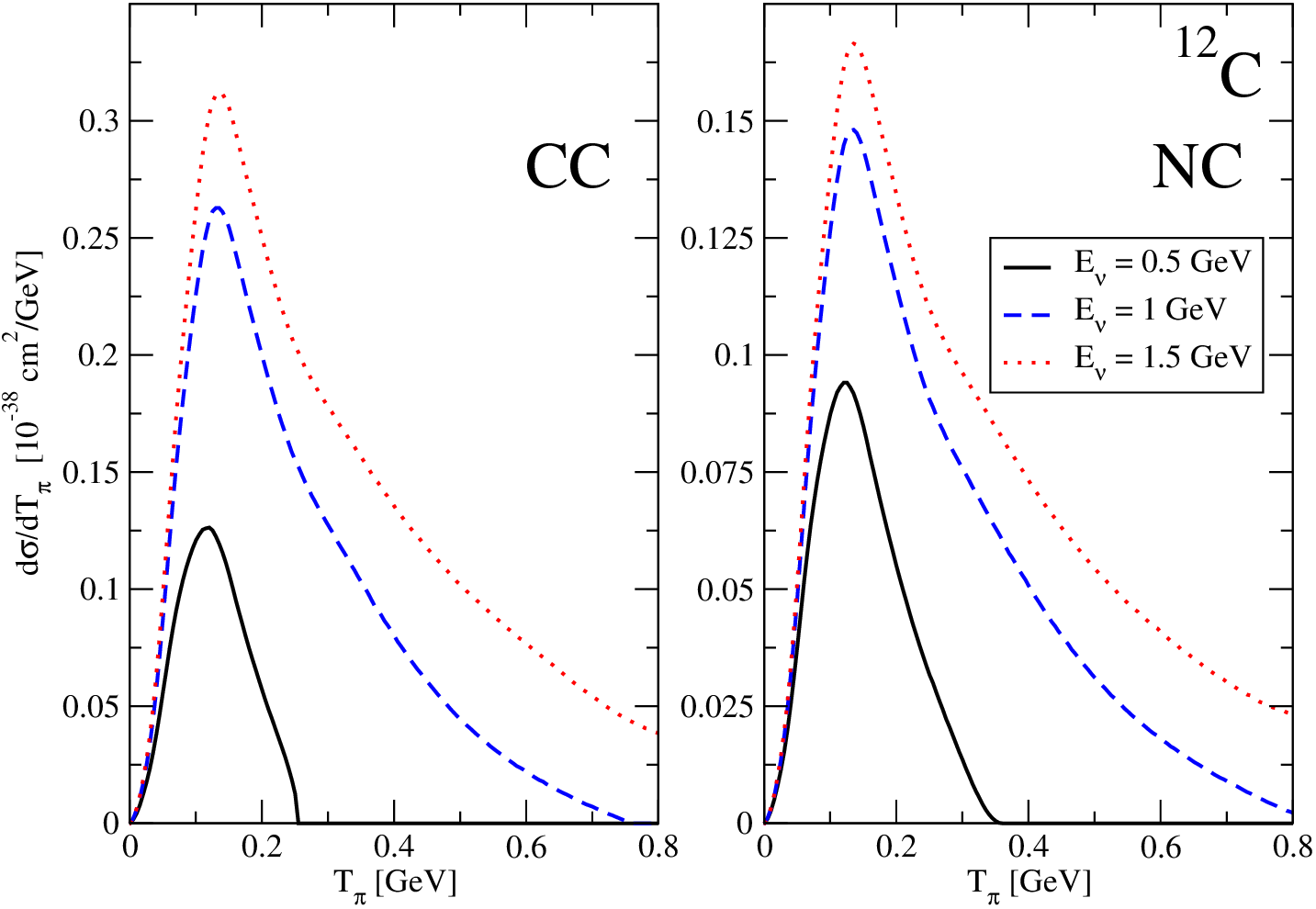}~~~~~~~~
\includegraphics[width=0.44\textwidth]{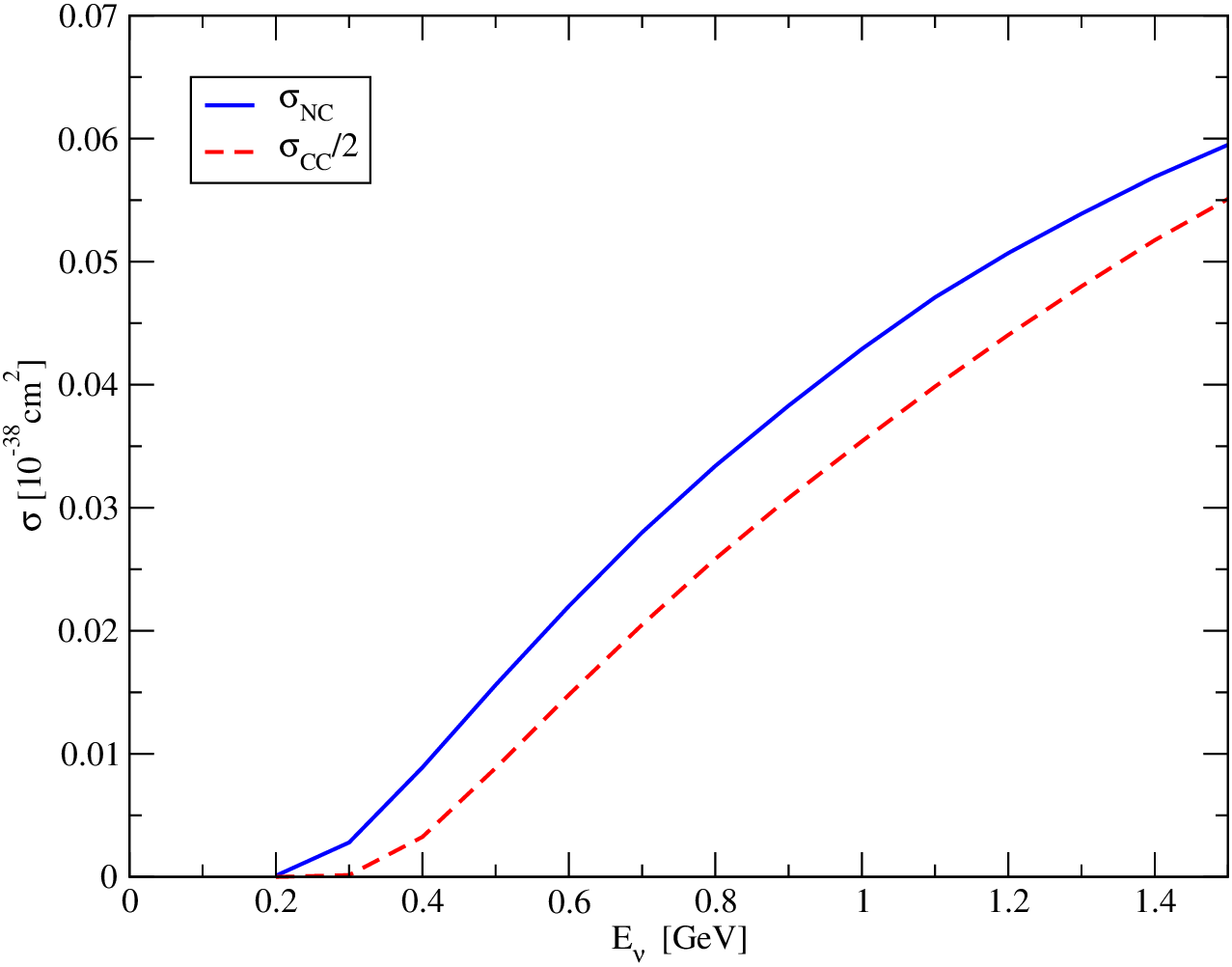}
\caption{Left (central) panel:  
differential cross-section for charged (neutral) current $\nu_{\mu}$-induced coherent pion 
production off $^{12}$C versus pion kinetic energy for several $\nu_{\mu}$ energies. 
Right panel: total cross-section for charged (divided by a factor 2) and neutral current 
coherent pion production as a function of the $\nu_{\mu}$ energy.}
\end{figure}
The response naturally associated to the coherent process is the 
spin-isospin longitudinal one, since it has the same coupling as the pion.
We have tested our description of the coherent responses on 
the elastic $\pi$ -$^{12}$C scattering which is sensitive to collective effects in the 
longitudinal channel. We have also checked the compatibility of our evaluation of 
neutrino differential cross-section in the forward direction with the experimental 
$\pi$ -$^{12}$C elastic scattering cross-section, according to Adler's theorem.

Figure \ref{fig_coh_diff} displays our evaluations of the coherent pion production off $^{12}$C 
as a function of the pion kinetic energy, both for charged and neutral current, 
for several neutrino incident energies. The resulting total coherent cross-sections 
are also shown as a function of the neutrino energy.

\subsection{Quasi-elastic and multi-nucleon channels}
 The quasi-elastic channel corresponds to a single-nucleon knock-out.
In contrast to the coherent channel, the quasi-elastic process 
is dominated by the transverse response.
The quasi-elastic cross-section is displayed in Fig.\ref{fig_qe_pi} 
as function of the energy transfer $\omega$, for a neutrino energy 
of 1 GeV,
both in the bare and in the RPA cases. 
The RPA effects tend to reduce the cross-section, 
as expected 
from the repulsive character of the particle-hole 
interaction which dominates in the transverse channel.  
In the same figure we display the sum of the two- and three-nucleon knock-out cross-section, 
which represents a sizable fraction of the quasi-elastic one. 
Singling out the genuine quasi-elastic process requires the insurance that no more than one proton is ejected. 
This issue will appear in connection with the comparison to data. 
Part of multi-nucleon channels arises from the modification of the Delta width 
in the medium where other decay channels are possible \cite{Oset:1987re}. 
The remaining contribution is taken from \cite{DELGUICH,Shimizu:1980kb}.
In neutrino interactions this last part of the cross-section is important but 
not very well constrained by phenomenology.

\subsection{Incoherent pion emission}
Turning now to incoherent pion emission,
the pion arises from the pionic decay of the Delta leaving the nucleus in a particle-hole excited state. 
For the nucleus that we consider, the incoherent pion cross-section is much larger than the coherent one.
 As compared to a free nucleon, the emission probability is already reduced 
in the bare case by the change in the Delta width. 
Moreover the RPA effects, which are moderate, also contribute to this reduction. 
The reduction due to the modification of the Delta width has a counterpart 
in the presence of the multi-nucleon knock-out component discussed before. 

\begin{figure}
\label{fig_qe_pi}
  \includegraphics[height=.27\textheight]{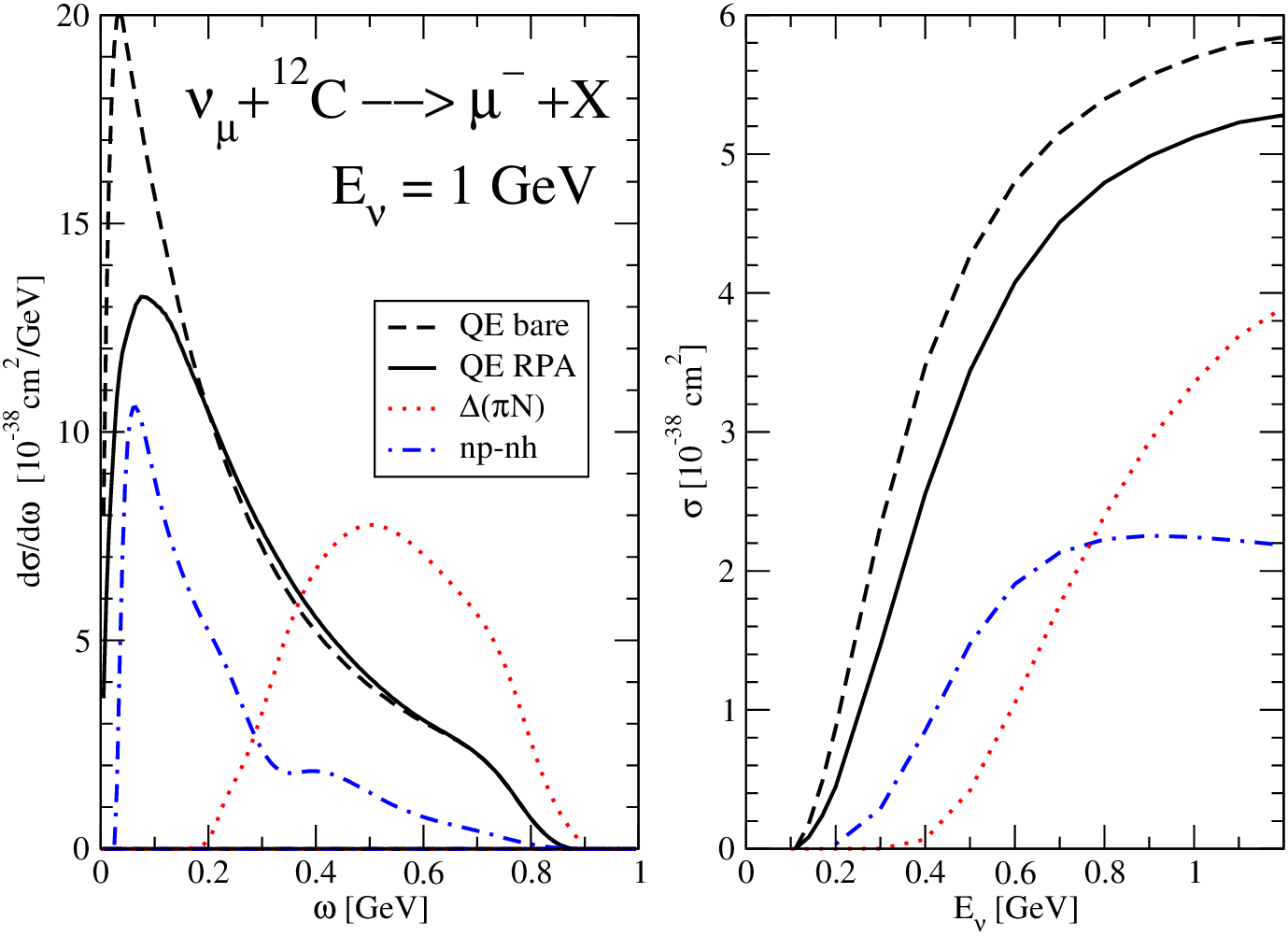}
  \caption{Left panel: differential $\nu_\mu$-$^{12}$C charged current 
cross-section as a function of transferred 
energy in the various channels for $E_\nu$=1 GeV. 
Right panel: total $\nu_\mu$-$^{12}$C charged current cross-section in the various channels.}
\end{figure}

All the previous results are summarized in Fig.\ref{fig_qe_pi} which displays the muon-neutrino 
differential cross-section in the various channels 
as a function of the energy transfer for the case of $^{12}$C
and a neutrino energy of 1 GeV. The total neutrino cross-section is also displayed. 
The incoherent $\pi$ channel includes all possible charge states. In our evaluation the 
incoherent $\pi^+$ channel results to be $5/6$ of the total.

\subsection{Comparison with data}

\begin{figure}
\label{fig_ratio_pip}
\includegraphics[width=0.45\textwidth]{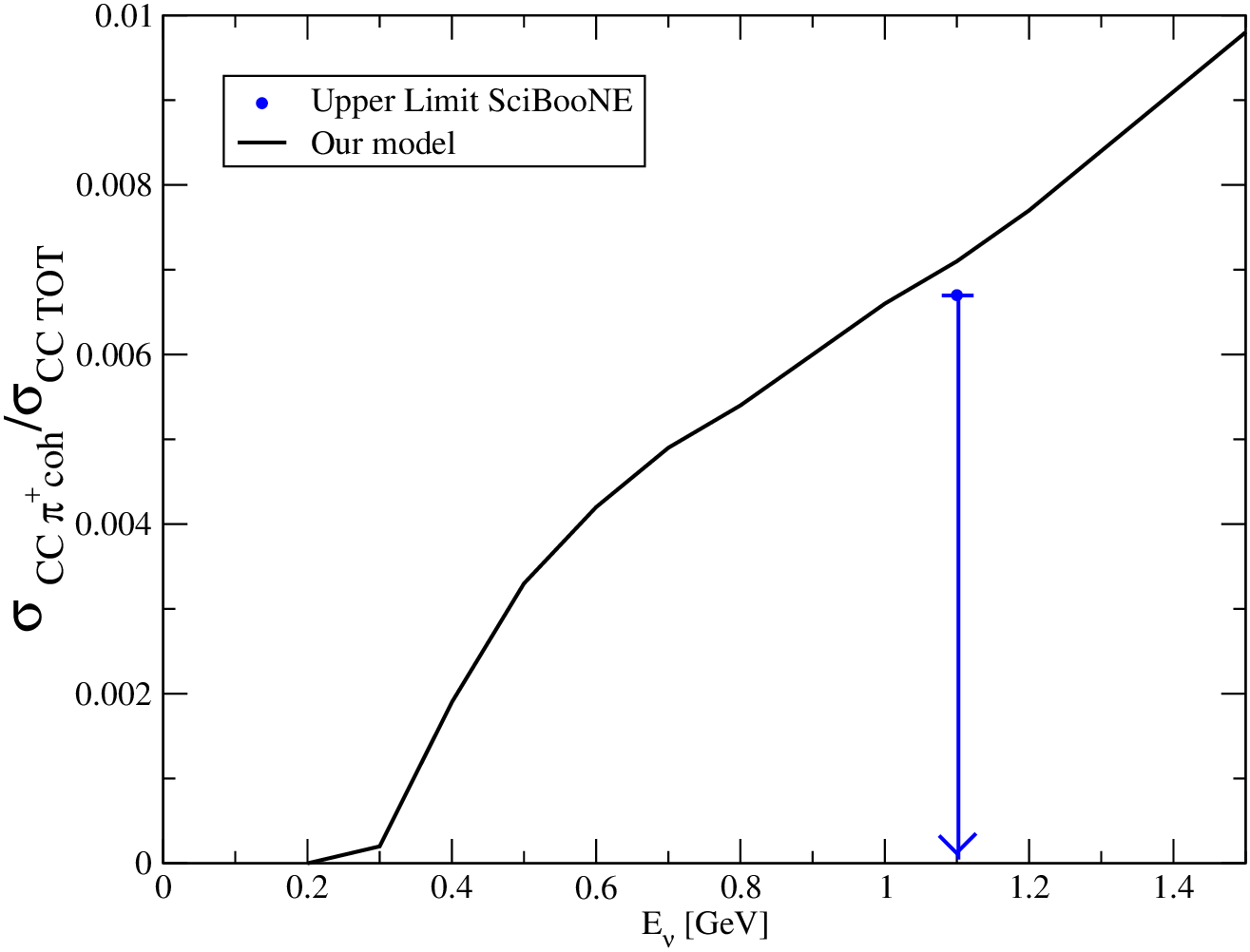}~~~~~~~~
\includegraphics[width=0.48\textwidth]{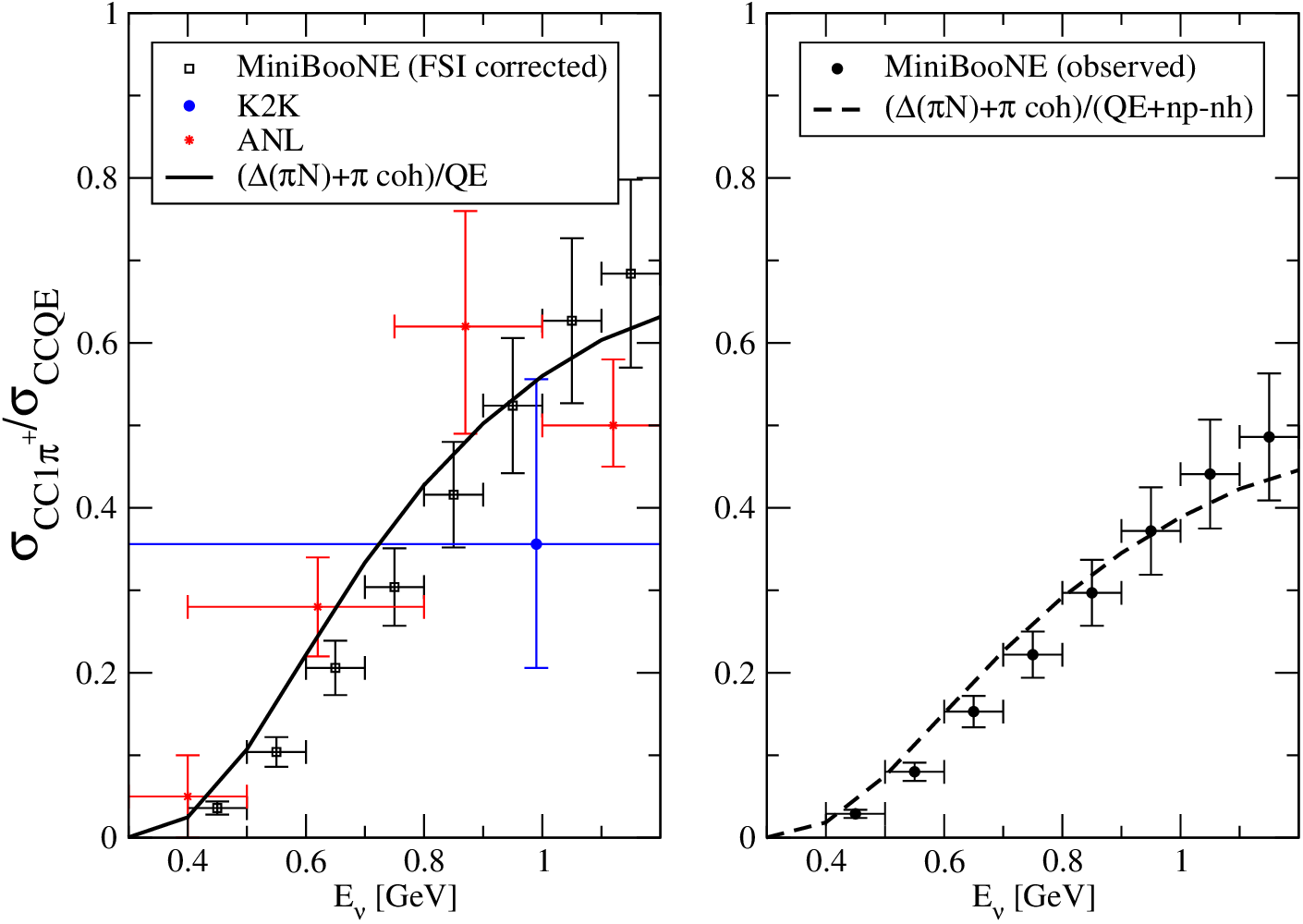}
\caption{Left panel: ratio of the $\nu_{\mu}$-induced charged current coherent
pion production to total cross-section. 
Central and right panels: ratio of the $\nu_{\mu}$-induced charged current one 
pion production to quasi-elastic cross-section.}
\end{figure}

Experimental data concern ratios between different cross-sections. 
For charged current, the K2K collaboration has established a 90\% 
confidence-level upper bound on the 
ratio of coherent pion production to 
the total cross-section, giving a 
limit of $0.60~10^{-2}$ averaged over the neutrino flux 
with a mean energy of 1.3 GeV \cite{Hasegawa:2005td}.
More recently, the SciBooNE collaboration found for the same quantity $0.67~10^{-2}$ 
at neutrino energy of 1.1 GeV \cite{Hiraide:2008eu}.
We report in the left panel of Fig.\ref{fig_ratio_pip} our prediction for this quantity. 
Our curve is just compatible with the experimental limit.

Another measured quantity is the ratio of $\pi^+$ production to 
quasi-elastic cross-section for charged current. The MiniBooNE collaboration has used a 
CH$_2$ target. In order to compare with ANL \cite{Radecky:1981fn} and K2K \cite{:2008eaa}
data, they presented the results applying an isoscalar rescaling correction \cite{AguilarArevalo:2009eb}. 
%In MiniBooNE a $CH_2$ target has been used but they have applied an isoscalar rescaling correction.
The issue of pion loss by final state interaction, which is not incorporated in our description, 
has also been taken into account by MiniBooNE 
who corrects data for this effect.
We can thus compare our $\pi^+$ over quasi-elastic ratio (solid line in the central 
panel of Fig.\ref{fig_ratio_pip})
to the final-state-interaction-corrected MiniBooNE results.
%The data points, as compiled by MiniBooNE, also include experimental results of \cite{Radecky:1981fn,:2008eaa}. 
Our curve is fully compatible with experimental data.

As an additional information, MiniBooNE also gives a ratio more directly related to the measurements,
namely the ratio of pion-like events 
(defined as events with exactly one $\mu^-$ and one $\pi^+$ escaping the struck nucleus) 
and quasi-elastic signal (defined as those with one $\mu^-$ and no pions). 
In our language the last quantity represents the $np-nh$ 
(including the quasi-elastic for $n=1$) exclusive channel. 
We have compared this second experimental information with 
the ratio between our calculated pion production (which however ignores final state interactions) 
and our total $np-nh$ 
contribution to the total charged current neutrino cross-section 
(right panel of Fig.\ref{fig_ratio_pip}). 
The comparison shows an 
agreement up to  $E_{\nu}\simeq 1.2$ GeV. This may be an indication that 
final state interactions for the pion is not 
essential here. Our theoretical approach predicts a strong 
contribution of $2p-2h$ and $3p-3h$ channels which 
seems to be supported by the last comparison.

%%%%%%%%%%%%%%%%%%%%%%%%%%%%%%%%%%%%%%%%%%%
%% The following lines show an example how to produce a bibliography
%% without the help of the BibTeX program. This could be used instead
%% of the above.
%%%%%%%%%%%%%%%%%%%%%%%%%%%%%%%%%%%%%%%%%%%

\end{document}